\documentclass[12pt]{article}
\usepackage[]{epsfig}
\usepackage[centertags]{amsmath}
\usepackage{amsfonts}
\usepackage{amssymb}
\usepackage[english]{babel}
\usepackage{amsmath, amsthm, slashed,url}
\usepackage{color}
\begin{document}

\title{\bf Spin-1/2 Landau levels in the symmetric gauge from the zero energy modes}
\author{ 
Lucas~Sourrouille$^a$,$^b$
\\
{\normalsize\it $^a$ IMDEA Nanociencia, Calle de Faraday, 9, Cantoblanco, 28049, Madrid, Spain}
\\
{\normalsize \it $^b$ IFISUR, Departamento de F\'isica (UNS-CONICET) }\\
{\normalsize\it  Avenida Alem 1253, Bah\'ia Blanca, 8000, Buenos Aires, Argentina }
\\
{\footnotesize  sourrou@df.uba.ar} } \maketitle

\abstract{Starting from the zero modes of the Dirac-Weyl equation for Landau levels in the symmetric gauge, we 
propose a novel mechanism to construct the eigenvalues and its eigenfunctions. We show that the problem may be addressed 
without 
numerical calculation and only solving the Dirac-Weyl equation for the zero modes. Specifically, the eigenstates associated to 
the negative magnetic field configurations may 
be constructed from the zero mode with positive chirality. In addition,  we obtain that the eigenstates associated 
to the 
positive magnetic field configurations may be constructed from the zero mode with negative chirality.
Finally, we show that our mechanism may be used to obtain the eigenvalues and eigenfunctions of the Hamiltonian 
corresponding to bilayer graphene system.}
 
\vspace{0.3cm}


\vspace{1cm}
\section{Introduction}
The problem of a single electron confined to two dimensions and exposed to a magnetic field was explored by Darwin \cite{D}, 
Fock \cite{F} and Landau \cite{L}. They show that the electron kinetic energy is quantized, being the discrete kinetic energy 
levels ``the Landau levels''.
\\
In particular, the Landau levels become relevant in the Quantum Hall problem \cite{1, 2, 3, 4, 5, 5.1, 5.2, 5.3, 5.4, 
5.5, 5.6}. 
Indeed, the integral Quantum Hall 
effect is a direct consequence of the Landau level formation. In addition the explanation for the fractional quantum Hall effect, 
arises because the lowest Landau level splits into Landau-like energy levels \cite{5.7, 5.8, 5.9, 5.10}. 
\\
In addition, deals with landau levels in planar geometry and in the symmetric gauge, acquire special significance, since 
the physics of the FQHE would not have revealed itself in a gauge other than the symmetric gauge of the planar geometry 
\cite{5.7, 
5.8}.
\\
On the other hand,
the experimental realization of monolayer graphene films \cite{1.1,2.1,3.1} has allowed explore the physics of 
two-dimensional (2D)
Dirac-Weyl fermions. This allows relativistic physics to be explored in a solid state system and physical phenomena such as the 
Klein-Gordon paradox, the anomalous Landau-Hall effect or nanoelectric materials \cite{4.1,5.1,3.1} may be addressed. Also, the 
study of Dirac-Wely electrons in magnetic fields has received much attention in the resent time in order to find a way for 
confining the charges \cite{6.1}-\cite{15.1}.
\\
In this note, we care to study the Landau levels for the Dirac-Weyl equation in the symmetric 
gauge. This problem was, previously, studied numerically in the Landau gauge (see for review see \cite{6}). In particular, in 
reference \cite{67}
exact analytical solutions for the bound states of the
Dirac-Wely electron in magnetic fields with various q-parameters under an electrostatic
potential were obtained. In addition, the Landau levels in the symmetric gauge were analyzed numerically for the nonrelativistic 
case \cite{677},\cite{6771}.
\\
Here, we show that the problem may be addressed without numerically calculation, and only solving the Dirac-Weyl equation for the 
zero modes. 
Specifically, we will develop a formalism
to construct the eigenvalues and its eigenfunctions of the Dirac-Weyl equation 
for the landau problem in the symmetric gauge.
This formalism is similar to the well know ladder operators mechanism, which is used to obtain the Landau 
levels for the Schr\"{o}dinger Hamiltonian \cite{6772}-\cite{6774} and allows us to generate all the eigenstates of the 
Hamiltonian 
from any one energy eigenstate. The novelty, here, is to develop a mechanism of ladder 
operators for Dirac-Weyl equation, which allows us obtain all Landau levels and their respective eigenfunctions. As we will see 
our formalism is not only applicable to Dirac-Weyl Hamiltonian, which govern the dynamics of monolayer graphene, but also to  
bilayer and multilayer graphene Hamiltonians.
Specifically,
we will show that, the eigenstates associated to the negative magnetic field 
configurations may 
be constructed from the zero mode with positive chirality. On the other hand, we obtain that the eigenstates 
associated to the 
positive magnetic field configurations may be constructed from the zero mode with negative chirality. 
In addition, we discuss how to generalize our mechanism to more complex problems such as the study of bilayer 
graphene system \cite{mc}-\cite{mm}.

\section{The framework and the Aharonov-Casher theorem}

Let us start by considering a $(2+1)$-dimensional Dirac-Weyl model whose Hamiltonian is described by
\begin{equation}
H= \sigma^i p_i = (\sigma^1 p_1 +  \sigma^2 p_2)\;,
\label{}
\end{equation}
Here, the $\sigma^i$ $(i =1,2)$ 
are 2$\times$2 Pauli matrices, i.e.
\begin{eqnarray}
\sigma^1 =\left( \begin{array}{cc}
0 & 1 \\
1 & 0 \end{array} \right)
\,,
\;\;\;\;\;\
\sigma^2 =\left( \begin{array}{cc}
0 & -i \\
i & 0 \end{array} \right)
\end{eqnarray}
and $p_i =-i\partial_i$ is the two-dimensional momentum operator. The massless Dirac-Weyl equation in $(2+1)$ dimensions is
\begin{equation}
\sigma^i p_i \Phi (x, y, t) = i\partial_t \Phi (x, y, t)
\label{eq1}
\end{equation}
Here, $\Phi (x, y, t)$ is the two-component spinor
\begin{equation}
\Phi=(\phi_a,\phi_b)^T
\label{}
\end{equation}
where $\phi_a$ and $\phi_b$ represent the envelope functions associated with the probability amplitudes. Since, we are interested 
in stationary states, it is natural to propose a 
solution of the form
\begin{eqnarray}
\Phi (x, y, t) = e^{-iEt} \Psi (x, y)\;,
\label{}
\end{eqnarray}
then, the time-independent Dirac-Weyl equation is
\begin{equation}
\sigma^i p_i \Psi (x, y) = E \Psi (x, y)
\label{1dw}
\end{equation}
In the presences of a perpendicular magnetic field to the $(x, y)$-plane, we replace the momentum operator $p_i$ by the covariant 
derivative, defined as $D_{i}= -i\partial_{i} +A_{i}$ $(i =1,2)$, where $A_{i}$ are components of the vector potential,
\begin{equation}
B=\partial_x A_y -\partial_y A_x
\label{mag}
\end{equation}
Thus, the equation (\ref{1dw}) becomes, 
\begin{equation}
\sigma^i D_i \Psi (x, y) = E \Psi (x, y)
\label{2dw}
\end{equation}
We can develop this equation to get,
\begin{equation}
\left( \begin{array}{cc}
0 & D_1 -iD_2 \\
D_1 +iD_2 & 0 \end{array} \right) \left( \begin{array}{c}
\psi_a\\
\psi_b \end{array} \right) = E \left( \begin{array}{c}
\psi_a\\
\psi_b \end{array} \right) 
\label{3dw}
\end{equation}
where $\psi_a$ and $\psi_b$ are the components of the spinor $\Psi$ (i.e. $\Psi=(\psi_a,\psi_b)^T$). From this equation we can
write the two coupled equations for the components $\psi_a$ and $\psi_b$
\begin{eqnarray}
D_1 \psi_b -iD_2 \psi_b = E \psi_a
\label{eqm1}
\end{eqnarray}
\begin{eqnarray}
D_1 \psi_a +iD_2 \psi_a = E \psi_b
\label{eqm2}
\end{eqnarray}
Here, we are interested to find the the eigenvalues and eigenstates corresponding to Eq.(\ref{2dw}). In order to find these one 
needs to specify a gauge for the vector potential. Here, we will use a symmetric gauge,
\begin{eqnarray}
{\bf A} = \frac{1}{2} (-By , Bx, 0)
\label{vec}
\end{eqnarray}
Then, equations (\ref{eqm1}) and (\ref{eqm2}) becomes, 
\begin{eqnarray}
\Big[[-i\partial_x -\partial_y] - z\frac{B}{2} \Big] \psi_b = E \psi_a
\label{eqm1.1}
\end{eqnarray}
\begin{eqnarray}
\Big[[-i\partial_x +\partial_y] - z^{\dagger}\frac{B}{2} \Big] \psi_a = E \psi_b
\label{eqm2.1}
\end{eqnarray}
where $z= ix + y$ and $z^{\dagger}= -ix + y$. The simplest solution of the equations (\ref{eqm1.1}) and (\ref{eqm2.1}) are the 
zero energy modes, that is the solutions for zero energy. This solution may be constructed explicitly following, the work done by 
Aharonov and Casher \cite{ahronov}. For this purpose we assume that the vector potential is divergenceless, which is clearly 
satisfied by (\ref{vec}). Then, one can introduce a scalar potential $\lambda(x, y)$ such that, 
\begin{eqnarray}
A_x = -\partial_y \lambda
\,,
\;\;\;\;\;\
A_y = \partial_x \lambda
\label{gau}
\end{eqnarray}
and due to the equation (\ref{mag}),
\begin{eqnarray}
B = \partial_x^2 \lambda + \partial_y^2 \lambda
\label{div}
\end{eqnarray}
Then, it is not difficult to find the solutions of the equations (\ref{eqm1.1}) and (\ref{eqm2.1}) for the energy zero case. 
Indeed, substituting in equation (\ref{eqm1.1})  
\begin{eqnarray}
\psi_b = f_b e^{-\lambda}
\end{eqnarray}
and setting $E=0$, we obtain,
\begin{eqnarray}
[-i\partial_x -\partial_y] f_b=0
\label{fb}
\end{eqnarray}
In similar way, if we propose 
\begin{eqnarray}
\psi_a = f_a e^{\lambda}
\end{eqnarray}
equation is reduced to 
\begin{eqnarray}
[-i\partial_x +\partial_y] f_a=0
\label{fa}
\end{eqnarray}
Thus, $f_a$ and $f_b$ are analytic and complex conjugated analytic entire functions of $z = ix + y$, respectively. 
\\
The equation (\ref{div}) has the following solution 
\begin{eqnarray}
\lambda ({\bf r}) = \int  d {\bf r}'G({\bf r}, {\bf r}') B({\bf r}')
\end{eqnarray}
where 
\begin{eqnarray}
G({\bf r}, {\bf r}')= \frac{1}{2\pi}\ln \Big(\frac{|{\bf r} - {\bf r}'|}{r_0}\Big)
\end{eqnarray}
is the Green function of the Laplace operator in two dimensions and $r_0$ is an arbitrary constant. 
According to Ref.\cite{ahronov} the magnetic flux $\Phi$ is localized in a restricted region so that for $r \to \infty$
\begin{eqnarray}
\lambda ({\bf r}) = \frac{\Phi}{2\pi} \ln \Big(\frac{r}{r_0}\Big)
\end{eqnarray}
and 
\begin{eqnarray}
\psi_{a,b} = f_{a,b} \Big(\frac{r}{r_0}\Big)^{\frac{\gamma \Phi}{2\pi}}
\label{sol}
\end{eqnarray}
where $\gamma =1$ and $-1$ for $\psi_a$ and $\psi_b$ respectively. Since the entire function $f(z)$ cannot go to zero in all 
directions at infinity, $\psi_{a,b}$ can be normalizable only assuming that $\gamma \Phi < 0$, that is, zero-energy solutions can 
exist only for one spin direction, depending on the sign of the total magnetic flux.
\\
Now, consider the case  $\Phi > 0$, then in view of (\ref{sol}) we have $\psi_a =0$ and 
\begin{eqnarray}
\psi_{b} =  f_{b} e^{-\lambda} \simeq f_{b} \Big(\frac{r}{r_0}\Big)^{\frac{- \Phi}{2\pi}}
\label{sol1}
\end{eqnarray}
The function $f_b$ is dictated by  (\ref{fb}) and it is not difficult to check that the solutions are polynomials of the form
\begin{eqnarray}
f_{b} =\sum_{i=0}^{j} a_i z^i
\end{eqnarray}
However, one can easily see from Eq.(\ref{sol1}) that the solution is integrable with the square only assuming that $j \leq N$ 
(we count $j$ from $j=0$), 
where $N$ is the integer part of $\frac{\Phi}{2\pi}$. For the case  $\Phi < 0$ we have
\begin{eqnarray}
\psi_{a} = f_{a} e^{\lambda} \simeq f_{a} \Big(\frac{r}{r_0}\Big)^{\frac{ \Phi}{2\pi}}\,,
\;\;\;\;\;\ \psi_b =0
\label{sol2}
\end{eqnarray}
where 
\begin{eqnarray}
f_{a} =\sum_{i=0}^{j} \tilde{a}_i (z^\dagger)^i
\end{eqnarray}
and
\begin{eqnarray}
j \leq N
\end{eqnarray}
Thus, the number of the independent states with zero energy for one spin projection is equal to $N+1$, and there are no 
such solutions for another spin projection. 

\section{The first level of energy and its eigenstates for negative magnetic field}
\label{4v}
Let us now concentrate on the construction of eigenstates corresponding to eigenvalues different form zero. To proceed we start 
by considering the simplest case,  
\begin{eqnarray}
\Psi_{0,0} =\left( \begin{array}{c}
\tilde{a}_0 e^{\lambda} \\
0  \end{array} \right)
\label{q+}
\end{eqnarray}
Here, the first subindex denote the number of independent state with zero energy and the second index denote level of energy. 
In other words the spinor (\ref{q+}) is the first independent zero energy mode. 
According to what we have seen (\ref{q+}) is a solution of the set 
\begin{eqnarray}
\Big[[-i\partial_x -\partial_y] - z\frac{B}{2} \Big] \psi_b = 0
\nonumber \\[3mm]
\Big[[-i\partial_x +\partial_y] - z^{\dagger}\frac{B}{2} \Big] \psi_a = 0
\label{}
\end{eqnarray}
Then, we can take the operator $\Big[[-i\partial_x -\partial_y] - z\frac{B}{2} \Big]$ and apply it to $\psi_a$,
\begin{eqnarray}
\Big[[-i\partial_x -\partial_y] - z\frac{B}{2} \Big] \psi_a = \tilde{a}_0[-i\partial_x  \lambda -\partial_y \lambda]e^\lambda 
-z\frac{B}{2}\tilde{a}_0 e^{\lambda}
\label{}
\end{eqnarray}
Thus, in view of (\ref{vec}), (\ref{gau}) and the definition of $z$, it not difficult to arrive to following result
\begin{eqnarray}
\Big[[-i\partial_x -\partial_y] - z\frac{B}{2} \Big] \psi_a =  
-z B \tilde{a}_0 e^{\lambda} =-z B \psi_a
\label{psi1}
\end{eqnarray}
To continue we can take the state $-z B \tilde{a}_0 e^{\lambda}$ and apply the operator $\Big[[-i\partial_x +\partial_y] - 
z^{\dagger}\frac{B}{2} \Big]$. Then, we have,
\begin{eqnarray}
\Big[[-i\partial_x +\partial_y] - 
z^{\dagger}\frac{B}{2} \Big](-z B \tilde{a}_0 e^{\lambda}) =  
-2B \tilde{a}_0 e^{\lambda} = -2B \psi_a
\label{psi2}
\end{eqnarray}
From, (\ref{psi1}), (\ref{psi2}) we conclude that
\begin{eqnarray}
 \Big[[-i\partial_x +\partial_y] - 
z^{\dagger}\frac{B}{2} \Big] \Big[[-i\partial_x -\partial_y] - z\frac{B}{2} \Big]\psi_a =  -2B \psi_a
\label{psi3}
\end{eqnarray}
This is crucial point since indicates that $\psi_a$ is an eigenstate of the operator $\Big[[-i\partial_x -\partial_y] - 
z\frac{B}{2} \Big] \Big[[-i\partial_x +\partial_y] - z^{\dagger}\frac{B}{2} \Big]$ with eigenvalue $-2B$. Thus, we can rename 
$-2B$ as $E^2$ so that 
\begin{eqnarray}
E = \pm \sqrt{-2B}
\label{}
\end{eqnarray}
Then we can think the equations (\ref{psi1}) and (\ref{psi2}) as 
\begin{eqnarray}
\Big[[-i\partial_x -\partial_y] - z\frac{B}{2} \Big] \psi_a =  E\; \psi_{a,0}
\label{ea0}
\end{eqnarray}
\begin{eqnarray}
\Big[[-i\partial_x +\partial_y] - 
z^{\dagger}\frac{B}{2} \Big] E\; \psi_{a,0} =  
E^2\;\ \psi_a = -2B \psi_a
\label{ea1}
\end{eqnarray}
where $\psi_{a,0}$ may be obtained dividing $(-z B \tilde{a}_0 e^{\lambda})$ by $E$,
\begin{eqnarray}
\psi_{a,0} = \frac{(-z B \tilde{a}_0 e^{\lambda})}{\pm \sqrt{-2B}} = \pm \sqrt{\frac{-B}{2}}z \psi_a 
\label{}
\end{eqnarray}
Therefore, dividing the equation (\ref{ea1}) by $E$, the equations (\ref{ea0}) and (\ref{ea1}) become 
\begin{eqnarray}
\Big[[-i\partial_x -\partial_y] - z\frac{B}{2} \Big] \psi_a =  E\; \psi_{a,0}
\label{}
\end{eqnarray}
\begin{eqnarray}
\Big[[-i\partial_x +\partial_y] - 
z^{\dagger}\frac{B}{2} \Big] \psi_{a,0} =  
E \psi_a 
\label{}
\end{eqnarray}
Comparing these last two equations with (\ref{eqm1.1}) and (\ref{eqm2.1}), we see that 
\begin{eqnarray}
\Psi_{0,1} =\left( \begin{array}{c}
\psi_{a,0} \\
 \psi_a \end{array} \right)
\end{eqnarray}
is an eigenstate of the Dirac-Weyl Hamiltonian with eigenvalue $\pm \sqrt{-2B}$. Here, it is important to point out that for the 
eigenvalue to make sense the magnetic field must be negative. This implies that $\Phi < 0$ which agrees with the result 
(\ref{sol2}).
\\
We can try to go even further and study the case in which $\psi_a$ is the first order polynomial in $z^\dagger$, that is 
$\psi_{a} = (\tilde{a}_0 + \tilde{a}_1 z^\dagger )e^{\lambda}$, then we have the following zero mode,
\begin{eqnarray}
\Psi_{1,0} =\left( \begin{array}{c}
(\tilde{a}_0 + \tilde{a}_1 z^\dagger )e^{\lambda} \\
0  \end{array} \right)
\end{eqnarray}
Again, we can take the operator $\Big[[-i\partial_x -\partial_y] - z\frac{B}{2} \Big]$ and apply it to $\psi_a$,
\begin{eqnarray}
\Big[[-i\partial_x -\partial_y] - z\frac{B}{2} \Big] \psi_a = -zB\Big[(\tilde{a}_0 + \tilde{a}_1 z^\dagger )e^{\lambda}\Big] 
-2\tilde{a}_1 e^{\lambda} = -zB\psi_a -2\tilde{a}_1 e^{\lambda}
\label{eq44}
\end{eqnarray}
Here we see that our result incorporates the term $-zB\Big[( \tilde{a}_1 z^\dagger )e^{\lambda}\Big] 
-2\tilde{a}_1 e^{\lambda}$ to the result obtained in (\ref{psi1}), of course, this term is due to the action of 
$\Big[[-i\partial_x -\partial_y] - z\frac{B}{2} \Big]$ on $\tilde{a}_1 z^\dagger e^{\lambda}$. Also, it is interesting to note 
that, writing in terms of $\psi_a$, the result of (\ref{psi1}) is $-zB\psi_a$ whereas in (\ref{eq44}) we obtain $-zB\psi_a 
-2\tilde{a}_1 e^{\lambda}$, that is, the result in  (\ref{eq44}) is not only reduced to $-zB\psi_a$, but also now appears the 
term $-2\tilde{a}_1 e^{\lambda}$.
\\
Now, we can try to apply  the operator $\Big[[-i\partial_x +\partial_y] - z^{\dagger}\frac{B}{2} \Big]$ to $-zB\psi_a 
-2\tilde{a}_1 e^{\lambda}$. This operation leads us to the following result,
\begin{eqnarray}
\Big[[-i\partial_x +\partial_y] - 
z^{\dagger}\frac{B}{2} \Big](-zB\psi_a 
-2\tilde{a}_1 e^{\lambda}) =  
-2B (\tilde{a}_0 + \tilde{a}_1 z^\dagger ) e^{\lambda} = -2B \psi_a
\label{eq45}
\end{eqnarray}
Again the result obtained in (\ref{psi2}) and (\ref{psi3}) is repeated. Therefore, $-2B$ is an eigenvalue of $\Big[[-i\partial_x 
-\partial_y] - z\frac{B}{2} \Big] \Big[[-i\partial_x +\partial_y] - z^{\dagger}\frac{B}{2} \Big]$ with eigenstate 
$\psi_a = (\tilde{a}_0 + \tilde{a}_1 z^\dagger ) e^{\lambda}$. Thus, the equations (\ref{eq44}) and (\ref{eq45}) may be rewritten 
as
\begin{eqnarray}
\Big[[-i\partial_x -\partial_y] - z\frac{B}{2} \Big] \psi_a = E\; \psi_{a,1}
\label{46}
\end{eqnarray}
\begin{eqnarray}
\Big[[-i\partial_x +\partial_y] - 
z^{\dagger}\frac{B}{2} \Big](E\; \psi_{a,1}) =  E^2\;\ \psi_a = -2B \psi_a\,,
\label{47}
\end{eqnarray}
with,
\begin{eqnarray}
E = \pm \sqrt{-2B}\,,
\label{}
\end{eqnarray}
and where we have renamed $-zB\psi_a -2\tilde{a}_1 e^{\lambda}$ as $E\; \psi_{a,1}$, which implies,
\begin{eqnarray}
\psi_{a,1}= \frac{-zB\psi_a -2\tilde{a}_1 e^{\lambda}}{\pm\sqrt{-2B}} =\pm\frac{1}{\sqrt{2}}[\sqrt{-B}z\psi_a -\frac{2\tilde{a}_1 
e^{\lambda}}{\sqrt{-B}}]
\label{}
\end{eqnarray}
So, the equations (\ref{46}) and (\ref{47}) reads as,
\begin{eqnarray}
\Big[[-i\partial_x -\partial_y] - z\frac{B}{2} \Big] \psi_a = E\; \psi_{a,1}
\label{}
\end{eqnarray}
\begin{eqnarray}
\Big[[-i\partial_x +\partial_y] - 
z^{\dagger}\frac{B}{2} \Big] \psi_{a,1} =  E\;\ \psi_a 
\label{}
\end{eqnarray}
Thus, we have constructed an new eigenstate  
\begin{eqnarray}
\Psi_{1,1} =\left( \begin{array}{c}
\psi_{a,1} \\
 \psi_a \end{array} \right)
\end{eqnarray}
of the Dirac-Weyl Hamiltonian with eigenvalue $\pm \sqrt{-2B}$.
\\
In order to obtained a general formula for the eigenstates of the Dirac-Weyl Hamiltonian, let us explore the case in which 
$\psi_a$ is the second order polynomial in $z^\dagger$, that is 
$\psi_a = (\tilde{a}_0 + \tilde{a}_1 z^\dagger + \tilde{a}_2 (z^\dagger)^2)e^{\lambda}$. Then, after a bit of algebra, we have 
\begin{eqnarray}
\Big[[-i\partial_x -\partial_y] - z\frac{B}{2} \Big] \psi_a &=& -zB\Big[(\tilde{a}_0 + \tilde{a}_1 z^\dagger + \tilde{a}_2 
(z^\dagger)^2)e^{\lambda}\Big] +
(-2\tilde{a}_1 -4\tilde{a}_2 z^\dagger )e^{\lambda} 
\nonumber \\
&=& -zB\psi_a + (-2\tilde{a}_1 -4\tilde{a}_2 z^\dagger )e^{\lambda}
\end{eqnarray}
Notice that the result is composed by the term $-zB\psi_a$ plus a polynomial of first order in $z^\dagger$. If we compare 
this result with the results of formulas (\ref{psi1}) and (\ref{eq44}), we will notice that the result to apply the operator
$\Big[[-i\partial_x -\partial_y] - z\frac{B}{2} \Big]$ to the state $\psi_a$ is composed by the term $-zB\psi_a$ plus a 
polynomial 
of a lower order in $z^\dagger$ than $\psi_a$.
\\
To proceed, we apply the operator $\Big[[-i\partial_x +\partial_y] -z^{\dagger}\frac{B}{2} \Big]$ to the state $-zB\psi_a + 
(-2\tilde{a}_1 -4\tilde{a}_2 z^\dagger )e^{\lambda}$. Thus, we arrive to 
\begin{eqnarray}
&&\Big[[-i\partial_x +\partial_y] - 
z^{\dagger}\frac{B}{2} \Big]\Big(-zB\psi_a + 
(-2\tilde{a}_1 -4\tilde{a}_2 z^\dagger )e^{\lambda}\Big)
\nonumber \\
&&= -2B \Big(\tilde{a}_0 + \tilde{a}_1 z^\dagger +  \tilde{a}_2 (z^\dagger)^2\Big) e^{\lambda} = -2B \psi_a
\label{}
\end{eqnarray}
Thereby, we can construct the state $\psi_{a,2}$, dividing $-zB\psi_a + (-2\tilde{a}_1 -4\tilde{a}_2 z^\dagger )e^{\lambda}$ by 
$\pm \sqrt{-2B}$,
\begin{eqnarray}
\psi_{a,2}= \frac{-zB\psi_a + (-2\tilde{a}_1 -4\tilde{a}_2 z^\dagger )e^{\lambda}}{\pm\sqrt{-2B}} 
=\pm\frac{1}{\sqrt{2}}[\sqrt{-B}z\psi_a +\frac{(-2\tilde{a}_1 -4\tilde{a}_2 z^\dagger )e^{\lambda}}{\sqrt{-B}}]
\label{}
\end{eqnarray}
and the eigenstate of the Dirac-Weyl Hamiltonian as 
\begin{eqnarray}
\Psi_{2,1} =\left( \begin{array}{c}
\psi_{a,2} \\
 \psi_a \end{array} \right)
\end{eqnarray}
\\
It is not difficult to generalize this idea to the case
\begin{eqnarray}
\psi_{a} = \sum_{i=0}^{j} \tilde{a}_i (z^\dagger)^i e^{\lambda}
\label{}
\end{eqnarray}
with
\begin{eqnarray}
j \leq N
\end{eqnarray}
The result is 
\begin{eqnarray}
\psi_{a,j}
=\pm\frac{1}{\sqrt{2}}\Big[\sqrt{-B}z\psi_a -\frac{1}{\sqrt{-B}} \sum_{i=0}^{j} 2i\; \tilde{a}_i\; (z^\dagger)^{i-1} 
e^{\lambda}\Big]
\label{aj}
\end{eqnarray}
Therefore, we can create an other new eigenstate of the Dirac-Weyl Hamiltonian, 
\begin{eqnarray}
\Psi_{j,1} =\left( \begin{array}{c}
\psi_{a,j} \\
 \psi_a \end{array} \right)
 \label{sp}
\end{eqnarray}
with eigenvalue $\pm \sqrt{-2B}$. When $j= N$ we have 
\begin{eqnarray}
\Psi_{N,1} =\left( \begin{array}{c}
\psi_{a,N} \\
 \psi_a \end{array} \right)
 \label{st1}
\end{eqnarray}
Here, it is important to remark that the state (\ref{st1}) is $(N+1)$-th eigenstate that we create and due to the Aharonov-Casher 
theorem  we can not create another independent eigenstate. 
So, for a level of energy $E=\sqrt{-2B}$ we can only have $(N+1)$-th independent eigenstates and in similar way for the level 
$E=-\sqrt{-2B}$ we have $(N+1)$-th independent eigenstates associated to it.
Another interesting aspect shows that the eigenstates $\Psi_{j,1}$ depend only on $\psi_a$ i.e. an eigenstate of positive 
chirality. Hence, it is not surprising that for the eigenvalue to make sense the magnetic field must be negative. This agrees 
with 
the result (\ref{sol2}) and involves a extension of the Aharonov-Casher result.

\section{The first level of energy and its eigenstates for positive magnetic field}
Let us consider how to construct eigenstates of the Dirac-Weyl Hamiltonian starting now by a simplest spinor of negative 
chirality, i.e.
\begin{eqnarray}
\Psi_{0,0} =\left( \begin{array}{c}
0 \\
a_0 e^{-\lambda}  \end{array} \right)
\label{q-}
\end{eqnarray}
as the spinor of (\ref{q+}) it is a zero mode of the Dira-Weyl Hamiltonian. The idea, now, is applicate the operator 
$\Big[[-i\partial_x +\partial_y] - z^{\dagger}\frac{B}{2} \Big]$ to $\psi_b$. This can be done easily and leads us to,
\begin{eqnarray}
\Big[[-i\partial_x +\partial_y] - z^{\dagger}\frac{B}{2} \Big] \psi_b = -a_0 B z^\dagger e^{-\lambda} = -z^\dagger B \psi_b
\label{-1}
\end{eqnarray}
Following the same idea we take $-z^\dagger B \psi_b$ and apply the operator $\Big[[-i\partial_x -\partial_y] - z\frac{B}{2} 
\Big]$,
\begin{eqnarray}
\Big[[-i\partial_x -\partial_y] - z\frac{B}{2} \Big] (-z^\dagger B \psi_b) =  2 a_0 B e^{-\lambda} = 2 B \psi_b
\label{-2}
\end{eqnarray}
Thus, if we compare with (\ref{psi2}) and (\ref{psi3}) we see that the Schr\"{o}dinger equation
\begin{eqnarray}
 \Big[[-i\partial_x -\partial_y] - z\frac{B}{2} \Big] \Big[[-i\partial_x +\partial_y] - 
z^{\dagger}\frac{B}{2} \Big] \psi_b =  2B \psi_b
\label{}
\end{eqnarray}
has now an eigenvalue with the same absolute value but different sign, i.e. $2B$. Therefore, we have that the energy, now reads
\begin{eqnarray}
E = \pm \sqrt{2B}\;,
\label{}
\end{eqnarray}
so that the equations (\ref{-1}) and (\ref{-2}), can be arranged to give, 
\begin{eqnarray}
\Big[[-i\partial_x +\partial_y] - 
z^{\dagger}\frac{B}{2} \Big]
 \psi_b =  E\; \psi_{b,0}
\label{}
\end{eqnarray}
\begin{eqnarray}
\Big[[-i\partial_x -\partial_y] - z\frac{B}{2} \Big] E\psi_{b,0} =  
E^2 \psi_b 
\label{p+p}
\end{eqnarray}
where 
\begin{eqnarray}
\psi_{b,0} = \frac{-z^\dagger B \psi_b}{\pm \sqrt{2B}} = \mp \sqrt{\frac{B}{2}} z^\dagger \psi_b 
\label{}
\end{eqnarray}
Hence, dividing  (\ref{p+p}) by $E$, we have, the set following set of equations
\begin{eqnarray}
\Big[[-i\partial_x +\partial_y] - 
z^{\dagger}\frac{B}{2} \Big]
 \psi_b =  E\; \psi_{b,1}
\label{eq71}
\end{eqnarray}
\begin{eqnarray}
\Big[[-i\partial_x -\partial_y] - z\frac{B}{2} \Big] \psi_{b,0} =  
E \psi_b 
\label{eq72}
\end{eqnarray}
We can compare this equations with (\ref{eqm1.1}) and (\ref{eqm2.1}). Then, it is clear that 
\begin{eqnarray}
\Psi_{0,1} =\left( \begin{array}{c}
\psi_{b} \\
 \psi_{b,0} \end{array} \right)
\end{eqnarray}
is an eigenstate of the Dirac-Weyl Hamiltonian with eigenvalue $\pm \sqrt{2B}$. In contrast with the previous section, here, the 
magnetic field must be positive, otherwise the eigenvalue becomes meaningless. This implies that $\Phi > 0$, and again, it agrees 
with the result of the Aharonov and Casher theorem.
\\
As in the previous section the idea may be generalized. Indeed, if we take  
\begin{eqnarray}
\psi_{b} = \sum_{i=0}^{j} a_i (z^\dagger)^i e^{-\lambda}
\label{74}
\end{eqnarray}
and follow the same procedure of the previous section, we can create a j-th eigenstate, with eigenvalue $E= \pm \sqrt{2B}$, of 
the Dirac-Weyl Hamiltonian
\begin{eqnarray}
\Psi_{j,1} =\left( \begin{array}{c}
\psi_{b} \\
 \psi_{b,j} \end{array} \right)
 \label{st}
\end{eqnarray}
where 
\begin{eqnarray}
\psi_{b,j}
=\mp\frac{1}{\sqrt{2}}\Big[\sqrt{B}z^\dagger\psi_b -\frac{1}{\sqrt{B}} \sum_{i=0}^{j} 2i\; a_i\; (z^\dagger)^{i-1} 
e^{-\lambda}\Big]
\label{}
\end{eqnarray}
Again, due to the Aharonov and Casher theorem, $j \leq N$, so that we can, only, create $N+1$ independent eigenstates associated 
to the eigenvalue $E = \sqrt{2B}$ and another $N+1$  associated to level $E = -\sqrt{2B}$

\section{The generalization for higher energy levels}
\label{4}
Let us concentrate in the generalization of the procedure studied in the previous section. We can start by considering the second 
level of energy. In order to proceed consider the equations (\ref{eq71}) and (\ref{eq72}) and repeat the procedure of the 
previous section, i.e. take the second component of the spinor 
\begin{eqnarray}
\Psi_{0,1} =\left( \begin{array}{c}
\psi_{b} \\
 \psi_{b,0} \end{array} \right)
\end{eqnarray}
and apply the operator 
$\Big[[-i\partial_x +\partial_y] - z^{\dagger}\frac{B}{2} \Big]$ to this state. Then, we have,
\begin{eqnarray}
\Big[[-i\partial_x +\partial_y] - z^{\dagger}\frac{B}{2} \Big] \psi_{b,0} = \mp \sqrt{\frac{B}{2}} \Big[[-i\partial_x 
+\partial_y] 
- z^{\dagger}\frac{B}{2} \Big] (z^\dagger \psi_b)
\label{}
\end{eqnarray}
Since, $[-i\partial_x +\partial_y] z^\dagger =0$ and 
\begin{eqnarray}
\Big[[-i\partial_x +\partial_y] - 
z^{\dagger}\frac{B}{2} \Big]
 \psi_b =  E_1\; \psi_{b,0}
\label{}
\end{eqnarray}
where $E_1 = \pm \sqrt{2B}$,
it is not difficult to arrive to
\begin{eqnarray}
\Big[[-i\partial_x +\partial_y] - z^{\dagger}\frac{B}{2} \Big] \psi_{b,0} = -B z^\dagger \psi_{b,0}
\label{80}
\end{eqnarray}
We can continue with the procedure and take the state $-B z^\dagger \psi_{b,0}$ and apply the operator  $\Big[[-i\partial_x 
-\partial_y] - z\frac{B}{2} \Big]$,
\begin{eqnarray}
\Big[[-i\partial_x -\partial_y] - z\frac{B}{2} \Big] (-B z^\dagger \psi_{b,0}) =  2 B  \psi_{b,0} - B z^\dagger 
\Big[[-i\partial_x 
-\partial_y] - z\frac{B}{2} \Big]\psi_{b,0} 
\label{+1+1}
\end{eqnarray}
Here, we can consider the formula (\ref{p+p}). Then, the term of (\ref{+1+1}) can be rewritten as 
\begin{eqnarray}
\Big[[-i\partial_x -\partial_y] - z\frac{B}{2} \Big] (-B z^\dagger \psi_{b,0}) =  2 B  \psi_{b,0} \mp B \sqrt{2B} z^\dagger
\psi_b 
\label{+2-}
\end{eqnarray}
The second term of (\ref{+2-}) may be arranged as 
\begin{eqnarray}
\mp B \sqrt{2B} z^\dagger
\psi_b = 2B \psi_{b,0} \;,
\label{}
\end{eqnarray}
so that 
\begin{eqnarray}
\Big[[-i\partial_x -\partial_y] - z\frac{B}{2} \Big] (-B z^\dagger \psi_{b,0}) =  4 B  \psi_{b,0} 
\label{84}
\end{eqnarray}
If we rename $4B$ as $E_2^2$, and create a new state
\begin{eqnarray}
\psi_{b,1} = \mp \frac{\sqrt{B}}{2} z^\dagger \psi_{b,0}\;,
\label{}
\end{eqnarray}
then, the equations (\ref{80}) and (\ref{84}) reads as 
\begin{eqnarray}
\Big[[-i\partial_x +\partial_y] - z^{\dagger}\frac{B}{2} \Big] \psi_{b,0} = E_2 \psi_{b,1}
\label{80.1}
\end{eqnarray}
\begin{eqnarray}
E_2 \Big[[-i\partial_x -\partial_y] - z\frac{B}{2} \Big] \psi_{b,1} =  E_2^2  \psi_{b,0} 
\label{84.1}
\end{eqnarray}
Dividing the last equation by $E_2$, it becomes clear that 
\begin{eqnarray}
\Psi_{0,2} =\left( \begin{array}{c}
\psi_{b,0} \\
 \psi_{b,1} \end{array} \right)
\end{eqnarray}
is an eigenstate of the Dirac-Weyl Hamiltonian with eigenvalue $\pm 2\sqrt{B}$. In general, we can get the spinor (\ref{st}). 
Since, it is a solution of the Dirac-Weyl equation, it satisfy 
\begin{eqnarray}
\Big[[-i\partial_x +\partial_y] - 
z^{\dagger}\frac{B}{2} \Big]
 \psi_b =  E_1\; \psi_{b,j}
\label{}
\end{eqnarray}
\begin{eqnarray}
\Big[[-i\partial_x -\partial_y] - z\frac{B}{2} \Big] \psi_{b,j} =  
E_1 \psi_b 
\label{}
\end{eqnarray}
We can repeat the previous steps to obtain  
\begin{eqnarray}
\Big[\bar{\partial} - z^{\dagger}\frac{B}{2} \Big] \psi_{b,j} &=& \mp \sqrt{\frac{B}{2}} \Big[ \bar{\partial}
- z^{\dagger}\frac{B}{2} \Big] (z^\dagger \psi_b )
\nonumber \\
&&\pm \frac{1}{\sqrt{2B}} \sum_{i=0}^{j} 2i\; a_i\; \Big[\bar{\partial}
- z^{\dagger}\frac{B}{2} \Big]\Big((z^\dagger)^{i-1} 
e^{-\lambda}\Big)
\label{8+2}
\end{eqnarray}
Here $\psi_b$ is dictated by the formula (\ref{74}) and $\bar{\partial} =[-i\partial_x +\partial_y]$. The formula (\ref{8+2}) may 
be developed easily if we consider the following equalities,
\begin{eqnarray}
&&\bar{\partial} z^\dagger =0
\nonumber \\[3mm]
&&\Big[ \bar{\partial}- 
z^{\dagger}\frac{B}{2} \Big]
\psi_b =  E_1\; \psi_{b,j}
\nonumber \\[3mm]
&&\Big[ \bar{\partial}- 
z^{\dagger}\frac{B}{2} \Big]e^{-\lambda} = -B z^\dagger e^{-\lambda}
\label{}
\end{eqnarray}
then, it is not difficult to show that,
\begin{eqnarray}
\Big[\bar{\partial} - z^{\dagger}\frac{B}{2} \Big] \psi_{b,j} = -B z^\dagger \psi_{b,j} \mp \sqrt{2B}  \sum_{i=0}^{j} i\; a_i 
(z^\dagger)^{i} e^{-\lambda}
\label{93}
\end{eqnarray}
Now, the application of the operator $\Big[\partial - z\frac{B}{2} \Big]$, where we have named $\partial=[-i\partial_x 
-\partial_y]$, on the state $-B z^\dagger \psi_{b,j} \mp \sqrt{2B}  \sum_{i=0}^{j} i\; a_i 
(z^\dagger)^{i} e^{-\lambda}$ leads us to 
\begin{eqnarray}
\Big[ \partial - z\frac{B}{2}\Big] \Big(-B z^\dagger \psi_{b,j} \mp \sqrt{2B}  \sum_{i=0}^{j} i\; a_i 
(z^\dagger)^{i} e^{-\lambda}\Big) &=& -B \Big[ \partial - z\frac{B}{2}\Big] (z^\dagger \psi_{b,j}) 
\nonumber \\
&&\mp \sqrt{2B}  
\sum_{i=0}^{j} i\; a_i 
\Big[ \partial - z\frac{B}{2}\Big] \Big ((z^\dagger)^{i} e^{-\lambda}\Big)\nonumber \\
\label{ww}
\end{eqnarray}
To proceed we use the following identities

\begin{eqnarray}
&&\partial z^\dagger = -2
\nonumber \\[3mm]
&&\Big[ \partial - z\frac{B}{2}\Big] \psi_{b,j}
= E_1\; \psi_b 
\nonumber \\[3mm]
&&\Big[ \partial - z\frac{B}{2}\Big] e^{-\lambda} = 0\;,
\label{}
\end{eqnarray}
then equation (\ref{ww}) may be developed to give
\begin{eqnarray}
\Big[ \partial - z\frac{B}{2}\Big] \Big(-B z^\dagger \psi_{b,j} \mp \sqrt{2B}  \sum_{i=0}^{j} i\; a_i 
(z^\dagger)^{i} e^{-\lambda}\Big) = 4 B \psi_{b,j}
\label{96}
\end{eqnarray}
The equations (\ref{93}) and (\ref{96}) are arranged, so we have
\begin{eqnarray}
\Big[\bar{\partial} - z^{\dagger}\frac{B}{2} \Big] \psi_{b,j} = E_2 \psi_{b,j+1} 
\label{}
\end{eqnarray}
\begin{eqnarray}
\Big[ \partial - z\frac{B}{2}\Big](E_2 \psi_{b,j+1}) = 4 B \psi_{b,j}
\label{96}
\end{eqnarray}
where, as in (\ref{80.1}) and (\ref{84.1}), $E_2 = \pm 2\sqrt{B}$ and the state $\psi_{b,j+1}$ is defined as
\begin{eqnarray}
\psi_{b,j+1} = \frac{1}{\pm 2\sqrt{B}}\Big(-B z^\dagger \psi_{b,j} \mp \sqrt{2B}  \sum_{i=0}^{j} i\; a_i 
(z^\dagger)^{i} e^{-\lambda}\Big)
\label{}
\end{eqnarray}
Thus, we have that the state,
\begin{eqnarray}
\Psi_{j,2} =\left( \begin{array}{c}
\psi_{b,j} \\
 \psi_{b,j+1} \end{array} \right)
\end{eqnarray}
is an eigenstate of the Dirac-Weyl Hamiltonian with eigenvalue $\pm 2\sqrt{B}$. In this way, we can obtain all independent 
state of 
the second Landau level of the energy $\pm 2\sqrt{B}$. Again, it is interested to note that there are $N+1$ independent states 
with energy $E_2 = 2\sqrt{B}$ and another $N+1$ associated to the energy $E_2 = -2\sqrt{B}$. We can repeat the procedure to 
create 
the eigenstates associated to the third levels of energy and so on. In addition, it is not difficult to imagine the same 
procedure for the case of eigenvalues and eigenstates associated to negative magnetic field configuration.
In this case, we would take the first component of the spinor (\ref{sp}) and would apply the operator $\Big[[-i\partial_x 
-\partial_y] - z\frac{B}{2} \Big]$, in order to obtain a state $\psi_{a,j+1}$ and a new spinor
\begin{eqnarray}
\Psi_{j,2} =\left( \begin{array}{c}
\psi_{a,j+1} \\
\psi_{a,j} 
 \end{array} \right)
\end{eqnarray}
which is the $j+1$-th eigenstate of the Dirac-Weyl Hamiltonian associated to the second level of energy $E_2 = \pm 2\sqrt{-B}$. 
Of course, as for the states associated positive magnetic field configuration, there are  $N+1$ independent states 
with energy $E_2 = 2\sqrt{-B}$ and another $N+1$ associated to the energy $E_2 = -2\sqrt{-B}$.

\section{Application to bilayer graphene}

To conclude, we discuss the application of our mechanism to study the case of bilayer Hamiltonian. The bilayer 
graphene \cite{mc}-\cite{mm} in the simplest approximation can
be considered as a zero-gap semiconductor with parabolic
touching of the electron and hole bands described by the
single-particle Hamiltonian.
By exfoliation of graphene one can obtain several layers of carbon atoms.
Its electronic
structure can be understood in the framework of a tight-binding model.
\\
Let us then consider 
the Hamiltonian for the bilayer graphene,
\begin{equation}
\left( \begin{array}{cc}
0 & (D_1 -iD_2)^2 \\
(D_1 +iD_2)^2 & 0 \end{array} \right) \left( \begin{array}{c}
\psi_a\\
\psi_b \end{array} \right) = E \left( \begin{array}{c}
\psi_a\\
\psi_b \end{array} \right) 
\label{3dw3}
\end{equation}
This description is accurate at the energy scale
larger than a few meV, otherwise a more complicated picture
including trigonal warping takes place; we will restrict our-
selves only by the case of not too small doping when the
approximate Hamiltonian (\ref{3dw3}) works. Two components of the
wave function originated from the crystallographic structure
of graphite sheets with two carbon atoms in the sheet per
elementary cell. There are two touching points per Brillouin
zone: K and K'. For smooth enough external potential, no
Umklapp processes between these points are allowed and
thus they can be considered independently.
\\ 
Thus, equation (\ref{3dw3}) may be rewritten as
\begin{eqnarray}
(D_1 -iD_2)^2 \psi_b = E \psi_a
\label{1eqm1}
\end{eqnarray}
\begin{eqnarray}
(D_1 +iD_2)^2 \psi_a = E \psi_b
\label{2eqm2}
\end{eqnarray}
Following the same procedure for the single-layer Hamiltonian we should find the eigenfunctions of the Hamiltonian (\ref{3dw3}) 
for zero energy in the symmetric gauge, that is,
\begin{eqnarray}
\Big[[-i\partial_x -\partial_y] - z\frac{B}{2} \Big]^2 \psi_b = 0
\label{1eqm1.1}
\end{eqnarray}
\begin{eqnarray}
\Big[[-i\partial_x +\partial_y] - z^{\dagger}\frac{B}{2} \Big]^2 \psi_a = 0
\label{2eqm2.1}
\end{eqnarray}
The solution of these equations can be found easily from the development done in section \ref{4v}. Indeed, we can check that 
\begin{eqnarray}
\psi_{b} =z^\dagger \sum_{i=0}^{j} a_i z^i e^{-\lambda}
\end{eqnarray}
and
\begin{eqnarray}
\psi_{a} = z \sum_{i=0}^{j} \tilde{a}_i (z^\dagger)^i e^{\lambda}
\end{eqnarray}
satisfy (\ref{1eqm1.1}) and (\ref{2eqm2.1}). In order to construct the eigenstates different from the zero modes we can start by 
considering the simplest solution in a negative field background. According to what we see in the previous sections, this 
solution should be a spinor 
\begin{eqnarray}
\Psi_{0,0} =\left( \begin{array}{c}
z \tilde{a}_0 e^{\lambda} \\
0  \end{array} \right)
\label{+q+}
\end{eqnarray}
which is the simplest zero mode of the Hamiltonian (\ref{3dw3}). Again, we can take the operator $\Big[[-i\partial_x -\partial_y] 
- z\frac{B}{2} \Big]$ and apply it to $\psi_a$. Thus, using the result of section \ref{4v} we obtain  
\begin{eqnarray}
\Big[[-i\partial_x -\partial_y] - z\frac{B}{2} \Big] \psi_a = 
-z^2 B \tilde{a}_0 e^{\lambda}
\label{}
\end{eqnarray}
Applying the operator $\Big[[-i\partial_x -\partial_y] - z\frac{B}{2} \Big]$ one more time, lead us to
\begin{eqnarray}
\Big[[-i\partial_x -\partial_y] - z\frac{B}{2} \Big]^2 \psi_a = 
z^3 B ^2\tilde{a}_0 e^{\lambda}
\label{w2}
\end{eqnarray}
Following the same reasoning carried out in the section \ref{4v}, we should apply the operator $\Big[[-i\partial_x +\partial_y] - 
z^{\dagger}\frac{B}{2} \Big]^2$ to the polynomial $z^3 B ^2\tilde{a}_0 e^{\lambda}$.  Again, using the results of the section 
section \ref{4v} and after a bit of algebra we arrive to
\begin{eqnarray}
\Big[[-i\partial_x +\partial_y] - 
z^{\dagger}\frac{B}{2} \Big](z^3 B^2 \tilde{a}_0 e^{\lambda}) =  
6 B^2 z^2 \tilde{a}_0 e^{\lambda} 
\label{}
\end{eqnarray}
The application of the operator $\Big[[-i\partial_x +\partial_y] -z^{\dagger}\frac{B}{2} \Big]^2$ for the second time lead us to 
following result
\begin{eqnarray}
\Big[[-i\partial_x +\partial_y] - 
z^{\dagger}\frac{B}{2} \Big](6 B^2 z^2 \tilde{a}_0 e^{\lambda}) =  
24 B^2 z \tilde{a}_0 e^{\lambda} = 24 B^2 \psi_a
\label{v2}
\end{eqnarray}
Renamed, $24 B^2$ as $E^2$ we have,
\begin{eqnarray}
E = \pm \sqrt{24} |B| 
\label{}
\end{eqnarray}
Then, the equations (\ref{w2}) and (\ref{v2}) may be rewritten as follow
\begin{eqnarray}
\Big[[-i\partial_x -\partial_y] - z\frac{B}{2} \Big]^2 \psi_a = 
E \psi_{a,0}
\label{2w2}
\end{eqnarray}
\begin{eqnarray}
\Big[[-i\partial_x +\partial_y] - 
z^{\dagger}\frac{B}{2} \Big]^2 \psi_{a,0} =  
E \psi_a
\label{2v2}
\end{eqnarray}
where $\psi_{a,0}$ is obtained by dividing $(z^3 B^2 \tilde{a}_0 e^{\lambda})$ by $E$,
\begin{eqnarray}
\psi_{a,0} = \frac{(z^3 B^2 \tilde{a}_0 e^{\lambda})}{\pm \sqrt{24} B} = \pm \frac{ z^3 B \tilde{a}_0 e^{\lambda}}{\sqrt{24}}
\label{}
\end{eqnarray}
This show that the spinor 
\begin{eqnarray}
\Psi_{0,1} =\left( \begin{array}{c}
\psi_{a,0} \\
 \psi_a \end{array} \right)
\end{eqnarray}
is an eigenstate of the Hamiltonian (\ref{3dw3}) with eigenvalue $\pm \sqrt{24} B$. In general, we can start by considering a 
zero mode of the form
\begin{eqnarray}
\Psi_{0,1} =\left( \begin{array}{c}
z \sum_{i=0}^{j} \tilde{a}_i (z^\dagger)^i e^{\lambda} \\
 0 \end{array} \right)
\end{eqnarray}
Then, we apply the operator $\Big[[-i\partial_x -\partial_y]- z\frac{B}{2} \Big]$ to $z \sum_{i=0}^{j} \tilde{a}_i 
(z^\dagger)^i e^{\lambda}$. Thus, in view of the equation (\ref{aj})  and $[-i\partial_x -\partial_y]z =0$ it not difficult to 
see that 
\begin{eqnarray}
\Big[[-i\partial_x -\partial_y]- z\frac{B}{2} \Big] z \sum_{i=0}^{j} \tilde{a}_i 
(z^\dagger)^i e^{\lambda} = \pm z \sqrt{-2B} \psi_{a,j}
\end{eqnarray}
where $\psi_{a,j}$ is dictated by the formula (\ref{aj}). The application of the operator $\Big[[-i\partial_x -\partial_y]- 
z\frac{B}{2} \Big]$ to $\pm z \sqrt{-2B} \psi_{a,j}$ leads to
\begin{eqnarray}
\Big[[-i\partial_x -\partial_y]- z\frac{B}{2} \Big]^2 z \sum_{i=0}^{j} \tilde{a}_i 
(z^\dagger)^i e^{\lambda} = \pm z \sqrt{-2B} \Big[[-i\partial_x -\partial_y]- z\frac{B}{2} \Big]\psi_{a,j}
\end{eqnarray}
where
\begin{eqnarray}
\Big[[-i\partial_x -\partial_y]- z\frac{B}{2} \Big]\psi_{a,j}= \pm\frac{1}{\sqrt{2}}\Big[|B|\sqrt{2}z \psi_{a,j} 
-\frac{1}{\sqrt{-B}} z^{-1} 2i (\pm \sqrt{-2 B})\psi_{a,j} \Big]
\end{eqnarray}
The same mechanism can be repeated starting from the zero mode 
\begin{eqnarray}
\Psi_{j,0} =\left( \begin{array}{c}
0 \\
\sum_{i=0}^{j} a_i z^i  e^{-\lambda}\end{array} \right)
\end{eqnarray}
In addition, following the same steps of the section \ref{4}, we can obtain the higher levels of energy and its eigenstates.
\\
The generalization of this formalism to a multilayer graphene, can be easily understood, from the form of the 
multilayer Hamiltonian equation \cite{gc1}-\cite{mm}
\begin{equation}
\left( \begin{array}{cc}
0 & (D_1 -iD_2)^J \\
(D_1 +iD_2)^J & 0 \end{array} \right) \left( \begin{array}{c}
\psi_a\\
\psi_b \end{array} \right) = E \left( \begin{array}{c}
\psi_a\\
\psi_b \end{array} \right) 
\label{3dw3}
\end{equation}
where $J$ is a positive integer. Then we can consider the solutions of the bilayer graphene and repeat the same formalism to 
obtain the solutions for a 3-layer graphene system and so on for an arbitrary number of layers. 

\section{Conclusion}

In summary, we have developed a formalism to construct the eigenvalues and its eigenfunctions of the Dirac-Weyl equation for the 
Landau problem in the symmetric gauge. We have shown that for negative magnetic field configurations the eigenstates may be 
constructed from the zero mode with positive chirality whereas for positive magnetic field the eigenstates are 
constructed form the zero mode with negative chirality. This is a consequence of the Aharonov-Casher theorem, which establishes 
that for a negative magnetic flux the zero-energy solutions can exist only for positive spin direction, whereas if the 
magnetic flux is positive the zero-energy solutions can exist only for negative spin direction. 
\\
In addition, we showed that our mechanism may be generalized to study the Hamiltonians of multilayer graphene systems, in such a 
way to obtain the eigenfunctions and eigenvalues for arbitrary numbers of graphene layers. This is important due to the role that 
Landau levels play in graphene.
The bilayer Hamiltonian is different both
from nonrelativistic (Schrodinger) and from relativistic (Dirac) cases. The eigenstates of this Hamiltonian have very special 
chiral properties \cite{gc1}, resulting in a special Landau quantization and special scattering. In this sense, our 
method becomes a useful tool in order to obtain the landau levels for bilayer graphene. Although, in this note we do not study in 
detail the bilayer graphene, it would be interesting to deal with more emphasis the study of Landau levels in bilayer and 
multilayer graphene. This becomes of particular importance due to the unconventional quantum Hall effect in bilayer graphene 
\cite{mc},\cite{gc1}, where Landau quantization of the fermions results in plateaus in Hall conductivity at standard integer 
positions, being missing the last (zero-level) plateau.

\vspace{0.6cm}

{\bf Acknowledgements}
\\
This work is supported by CONICET.

\end{document}